\documentclass[aps,prd,nofootinbib,twocolumn,longbibliography]{revtex4-1}
\usepackage[english]{babel}
\usepackage[autostyle, english = british]{csquotes}
\usepackage[CJKbookmarks, pdftex, bookmarksnumbered, bookmarksopen, colorlinks, citecolor=blue, linkcolor=blue]{hyperref}
\usepackage{amsmath,amssymb}
\usepackage{graphicx}
\usepackage{enumitem}
\setlist[enumerate]{topsep=0pt,parsep=-1mm,leftmargin=5mm,}
\def\be{\begin{equation}}
\def\ee{\end{equation}}

\begin{document}

\title{\large The thermodynamic cost of choosing}

\author{Carlo Rovelli}
\affiliation{Aix-Marseille University, Universit\'e de Toulon, CPT-CNRS, F-13288 Marseille, France.}\affiliation{Department of Philosophy and Rotman Institute, 1151 Richmond St.N London, ON   N6A5B7}\affiliation{Perimeter Institute, 31 Caroline Street N, Waterloo ON, N2L2Y5, Canada}

\begin{abstract} 
\noindent Choice can be defined in thermodynamical terms, and shown to have a thermodynamic cost: choosing between a binary alternative at temperature $T$ dissipates an energy $E\ge kT\ln 2$. 

\end{abstract}
\maketitle

In a celebrated paper, Landauer noticed that {\em erasing information} requires dissipation \cite{Landauer}.  The precise status of this observation is controversial (see a discussion and extended references in \cite{Norton}), but the idea has had a large impact on our understanding of the thermodynamics involved in computation and information.  Here I observe that there is a related connection between {\em choosing} (which {\em generates } information, instead of erasing it) and dissipation. This statement can be given a precise meaning under a definition of choice, discussed below.   With this definition, a choice has a thermodynamic cost: choosing between a binary alternative at temperature $T$ dissipates an energy $E\ge kT\ln 2$, where $k$ is the Boltzmann constant.

The verb ``to choose" is used in a variety of contexts.  One might say for instance that a thermostat ``has chosen" to turn the heating on. This is not the kind of choice that here I am interested in, because a thermostat  functions on the basis of a predictable mechanism. Here I rather reserve the term ``choice" for a process where {\em the outcome of the choice is stable (on a given time scale) and is not predictable}. This definition includes, but is much wider than, human's choices, which are specifically characterized by far more complex mechanisms in addition to the simple thermodynamical ones considered here (on this, see for instance \cite{Ismael}).   

I consider the classical context first, then the quantum one.  In the context of classical mechanics, lack of predictability is necessarily associated with incomplete information.  Therefore when when we talk about choice we are always talking about our ignorance of an underlying dynamical process determining the outcome \cite{Spinoza}.  

We can formalize this situation by considering a system $S$ with a phase space $\Gamma$ and an incomplete set of $N$ variables $A_n:\Gamma\to I\!\!R$, with $n=1,...,N$, that capture our incomplete knowledge.  Here `incomplete' means that the map into $I\!\!R^n$ defined by the set of the $A_n$'s is not injective. 

This structure defines a thermodynamic-like context, because we can distinguish  `micro-states' $s\in\Gamma$, from `macro-states' $a_n$ that are the possible values that the observables $A_n$ can take.   Each macro-state $a_n$ defines a region $R(a_n)$ in $\Gamma$ as its inverse image $R(a_n)=\{s\in \Gamma, A_n(s)=a_n\}$. If the phase space has a natural associated volume form $V$,  each macro-states has an associated entropy $S(a_n)$ defined by $S=k\ln V(R(a_n))$.  The entropy of a macro-state measures how many micro-states it contains, and therefore the amount of ignorance about the details, if we only the macro-state is known.  

\begin{figure}[t]
\includegraphics[width=5cm]{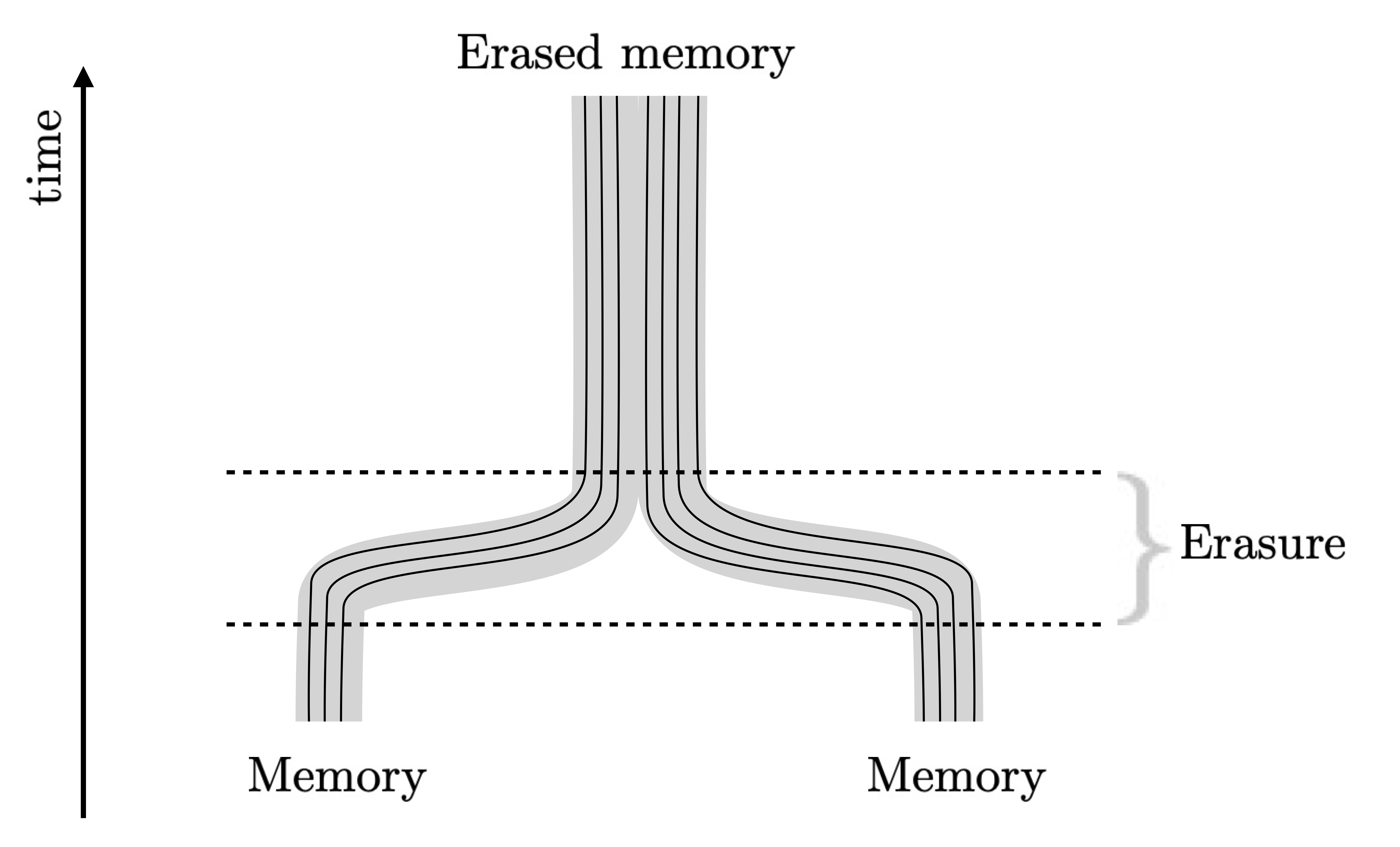}
\caption{Landauer memory erasure. Black lines represent micro-histories; thick grey lines macro-histories. The system is initially in one of two macro-states; time merges the two..}
\label{1}
\end{figure}

A microscopic history is described by a function $s(t)$, where $t$ is time. A macroscopic history is described by the functions $a_n(t)$.  A  macro-history $a_n(t)$ defines an ensemble of all micro-histories $s(t)$: those for which $A_n(s(t))=a_n(t)$.   We say that a macro-history is \emph{dissipative} where its entropy increases: $dS(a_n(t))/dt>0$.

 Because of the determinism of classical mechanics, knowledge of $s(t)$ for $t<0$ is sufficient to determine  $s(t) $ uniquely also for $t>0$. But not so for macroscopic histories.  That is, it is possible to have two macroscopic histories $a_n(t)$ and $a'_n(t)$, both compatible with the dynamics of the system such that 
\begin{eqnarray}
a_n(t)&\ne&a'_n(t),\ {\rm for}\ t<-\epsilon<0,\nonumber \\
a_n(t)&=&a'_n(t),\ {\rm for}\ t>\epsilon>0
\label{e}
\end{eqnarray}
or 
\begin{eqnarray}
a_n(t)&=&a'_n(t),\ {\rm for}\ t<-\epsilon<0,\nonumber \\
a_n(t)&\ne&a'_n(t),\ {\rm for}\ t>\epsilon>0
\label{choice}
\end{eqnarray}

\begin{figure}[b]
\includegraphics[width=5cm]{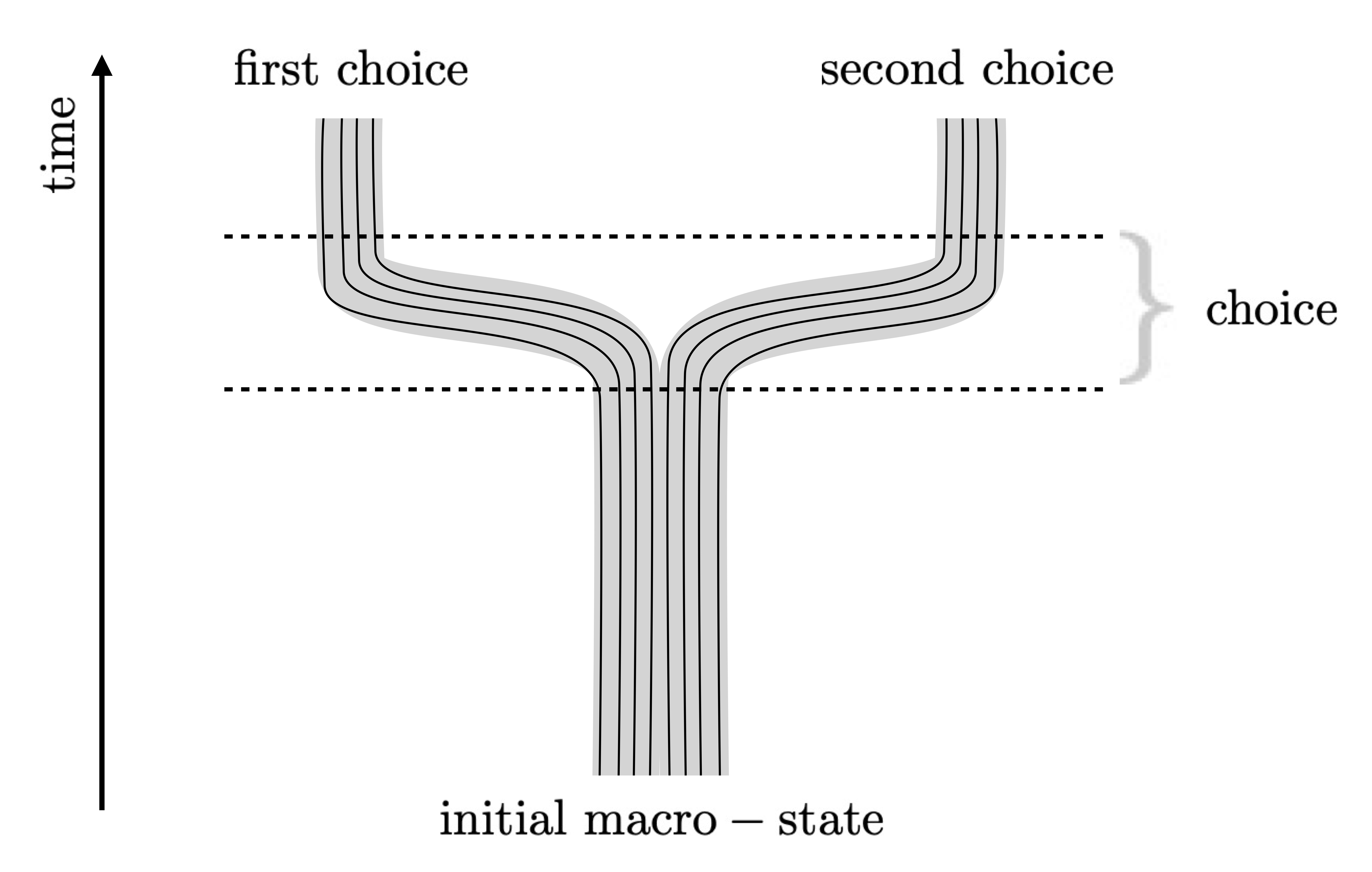}
\caption{A choice. A macro-state splits into two macro-states, each carrying a fraction of the micro-states.}
\label{1b}
\end{figure}

These two processes are represented by large grey branches in Figures \ref{1} and \ref{1b}, respectively. The first of these processes (Figure \ref{1}) can be interpreted as representing a Landauer memory erasure: erasure is a process that merges two macro-histories.   Information is not lost at the micro-level, where evolution is deterministic.   But information is lost is at the macro-level. That is, if we follow one of the branches we notice that the number of micro-states, hence the entropy, increases, By definition, there is dissipation.  This can be seen as minimal conceptual realization of a Landauer erasure. Say that there are $N_1>0$ micro-histories that start in the left macro-history and $N_2>0$ micro-histories that start in the right macro-history.    Say that the world is in one of the micro-histories, for instance one that starts in the left macro-state.  Before the erasure, the entropy of the actual macro-state of the world is $S_i=k\ln{N_1}$; after the erasure, the entropy of the actual macro-state of the world is $S_f=k\ln{(N_1+N_2)}>S_i$. Hence erasure increases entropy.  If $N_1=N_2=1$, the increase of entropy is $\Delta S = k\ln2$.  This is a form of the Landauer principle, that quantifies the minimal cost for erasing a memory.\footnote{Recall that I am in the context of a closed system, with no thermal bath.}

 In the second process (Figure \ref{1b}), the macroscopic system chooses between two different alternatives, both compatible with the dynamics, during the interval $[-\epsilon,\epsilon]$.  This is what happens when we make a choice by  flipping a coin: the macroscopic future is determined by a small fluctuation in the air's density, that determines whether the coin ends up  head  or cross.  The micro-history is deterministic, but the macro-history is not.    


The simple argument above might lead one to expect that a choice {\em decreases} entropy.  This is {\em not} the case, and this is the main point I make in this article. 

The reason the entropy does not actually decrease in a choice is that something needs to \emph{stabilize} the information transferred from the microphysics to the macro-physics by the choice. This differentiates the choice from the memory erasure.  In the erasure, the entropy grows, therefore the standard irreversibility implied by equilibration is sufficient to stabilize the erasure. Intuitively: the micro-state migrates to a larger region of phase space and is statistically unlikely to come back.   In a choice, described by the second figure, on the other hand, the entropy appears to decrease, and therefore equilibration tends to take it back immediately.   The two macroscopic states in which the process in Figure \ref{1b} ends up are immediately merged again by equilibration unless something stabilizes their outcome.   In the absence of something that stabilizes the choice, what Figure \ref{1b} describes is the first phase of a thermal fluctuation.  The flipped coin needs to hit the ground and dissipate its kinetic energy in order to providing us with what we call a choice.   Let analyse this phenomenon in detail in the context of an example.

\begin{figure}[t]
\includegraphics[width=3cm]{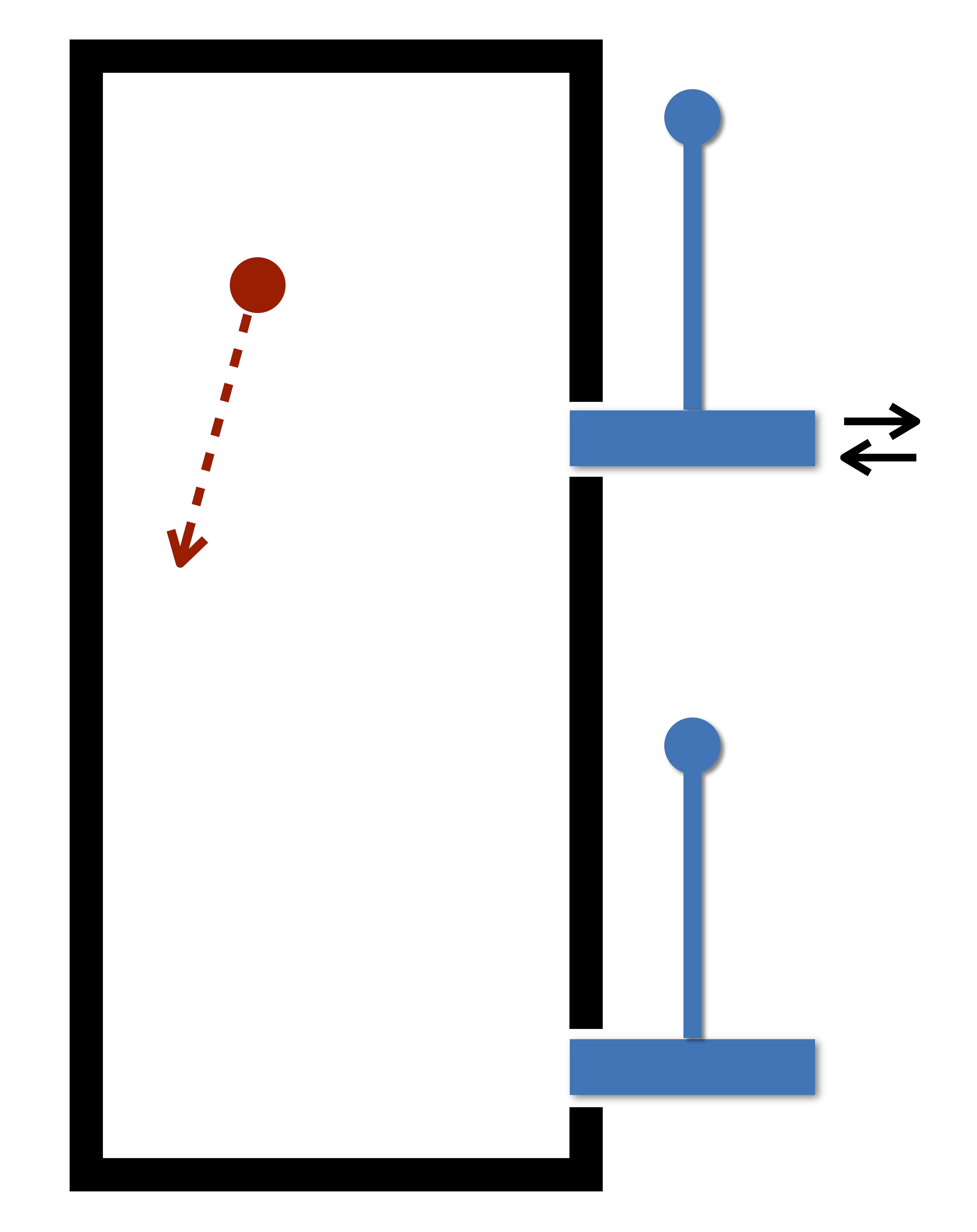}
\caption{A mechanical choice device. The ball, whose exact position is not known, hits first one or the other of the two penduli.}
\label{2}
\end{figure}

Consider the device depicted in Figure \ref{2}. It is formed by a single particle moving fast in a cavity with two small windows.  At each window there is a pendulum.  If the ball hits a pendulum, it may set it in motion. If we take the position and velocity of the particle to be micro-variables and the energy in the pendulums to be macro-variables, this can be considered as a devise that makes a choice.  In the macro-world, that is, evolution is unpredictable: it is determined by the unknown micro-history: if the pendulums are initially at rest,  which one of the two is set in motion first is determined by the actual position and velocity of the particle, which play here the role of the flipping coin.  

Given the distinction between micro and macro variables that we have stated, we can (unconventionally, as there is no thermal bath here) use a thermodynamical language, in spite of the small number of degrees of freedom.

We associate a temperature $T_h$ to the particle (which has three degrees of freedom), by $E=3\times\frac12 k T_h$, where $E$ is the energy of the particle, and a temperature $T$ to each pendulum, defined by $\epsilon=\frac12 k T_h$, where $\epsilon$ is the energy of the pendulum.  If the pendulums are initially exactly or nearly at rest, they have zero or low  energy, hence, according to our definition, low temperature. A collision transfers a bit of energy from the particle to one pendulum, namely from the higher temperature system to the lower temperature system. In thermodynamical terms, this implies an increase of entropy. This matches with the definition of entropy given above, because a step towards energy equipartition increases the accessible phase space region.   As long as $T_h>T$ it is `more probable' (in this sense) that in the collision energy is transferred from the particle to the pendulum.  The first pendulum hit receives energy and registers the choice in the macro-world.  For this to happen, the transferred $\Delta E$ energy must be larger than the thermal energy of the pendulum $\epsilon=\frac12 k T$, otherwise the oscillation of the pendulum caused by the collision cannot be distinguished from the thermal fluctuations. The increase in entropy due to the transfer of energy from a hot to a cold system is 
\be
\Delta S= \frac{\Delta E}{T}- \frac{\Delta E}{T+\Delta T}
\ee
where, as usual,  the two terms refer to the change of entropy of the particle and the pendulum, respectively, and $T_h=T+\Delta T$. The increase of information due to the outcome of the binary choice is a single bit.  For this to be consistent with the second principle, the overall entropy increase must at least compensate it, namely $\Delta S> k \ln 2$, giving 
\be
\Delta E \left(\frac{1}{T}- \frac{1}{T+\Delta T}\right)   >k \ln 2
\ee
By taking $\Delta T$ arbitrary large we can minimize the parenthesis, leaving 
\be
\Delta E   >k T \ln 2,
\label{bound}
\ee
which is the relation we were seeking.  That is, the minimal energy that must be dissipated in the binary choice at temperature $T$ is $\Delta E=k T \ln 2$.  

This mechanism is general. In the example, the stabilization of the choice is due to a definite increase in entropy, namely the transit of the microstate into a much larger region of phase space. This transit is implemented by the transfer of energy from the particle to the pendulum. Since we had assumed that the pendulum had a lower temperature (energy), the transfer of energy from the particle to the pendulum moves the overall system  towards energy equipartition and therefore into a much larger region of phase space, from which coming back is unlikely.  In general, in the statistical context considered, the {\em only} mechanism that can achieve a stabilization is clearly this transfer to a much larger region of state space, namely increase in entropy, namely dissipation, because without it, nothing prevents the system from (ergodically) rapidly fluctuate back and undo the choice.   In condensed words, stabilization must be irreversible, otherwise it is not stabilization.   In order to be irreversible a phenomenon must raise entropy, otherwise it is reversible. 

In \cite{Szilard}, Szilard pointed out that any measurement requires dissipation.  The same observation  was made earlier by Reichenbach \cite{Reichenbach}.  The argument above shows that any choice (as defined above) requires a ``measurement" in the sense of Szilard or Reichenbach. Specifically, Szilard derived a bound on the entropy produced in this process (Eq.(1) in \cite{Szilard}).  If we apply Szilard's analysis to the mechanical choice device of Fig. 3, Szilard's Eq.(1)  simplifies to $2ke^{-\Delta S} \ge1$ which is equivalent to \eqref{bound}.\footnote{I thank an anonymous referee for pointing out these connection.}

The regime considered here and the general framework, on the other hand is not the one considered by Szilard. There is no external thermal bath, no intervention on the system, no partition of the cavity inserted or extracted.  The reason is that we cannot understand the physics underpinning choices within an interventionist approach where agency, hence the possibility of choices, is {\em assumed}. We cannot assume what we want to understand.  Rather, we must look for phenomena that can be interpreted as choices and which happen with closed physical systems.  This cannot be done at equilibrium, because there is no dissipation at equilibrium.  The minimal framework considered is sufficient to make sense of this kind of dissipative phenomena (see also \cite{Rovelli2020,Rovelli2020a,Rovelli1956}), without need to resort to the full power of stochastic thermodynamics.   The key is that we are looking at a dynamical process where the definition of macroscopic variables is sufficient to define entropy, but we consider time intervals shorter than the thermalization times, so that the choice can be recorded can exist \cite{Rovelli2020}. 
In this regime, it is meaningful to identify the choice with `which pendulum is hit first', and the system has no time to equilibrate because the kinetic energy of the particle remains higher than that of the pendulums during the time considered,. The dissipation considered is not into a thermal bath: it is the transfer of energy from the higher energy particle to the lower energy pendulum.    Notice that in this language, ``stabilization" is meaningful only for times that are not too long, compared to thermalization times. On the long run, everything thermalizes, macro-states get to equilibrium, and no choice is made.   This notion of stabilization is coherent with what we usually call ``choice" and that it allows us to better understand the actual macroscopic world around us.

Choosing appears to be a phenomenon that is in a sense the inverse of erasing information (because branching is the inverse of merging). Therefore one may be puzzled by the above result and rather be tempted to expect that if erasing macroscopic information raises thermodynamical entropy, then choosing would decrease it.   The reason this is wrong is that there is no reversibility in dissipative phenomena.  A cyclic sequence  that produces a choice and then erases it repeatedly is not reversible: it continuously increase entropy at every step, like most real macroscopic phenomena. Entropy is raised both in choosing and in erasing the choice. The  information (negative-entropy) gained by the split of the macroscopic state must be compensate by a larger entropy increase (diffusion, energy equipartition,...), in order for the process to happen and stabilize. 

Finally, a few words about quantum theory.  At first sight, it seems that quantum theory can give us a free ride, circumventing the result above.  For instance, a Stern-Gerlach apparatus can give rise to a binary choice.  But a moment of reflection shows that this is not the case: we cannot disregard the physics of the apparatus and the environment.  This cannot be disregarded  if we are concerned with the thermodynamical balance. 

In fact, as nicely pointed out in \cite{Max}, the conventional textbook description of a quantum measurement appears to violate all three laws of thermodynamics, if we disregard what happens to apparatus and envoronment!   Thermodynamical legality is restored by considering the thermal properties of the apparatus and the environment, where entropy raises in order to store the result of the measurement  into the `classical' macro-physics.  See  \cite{Max} for a detailed discussion of this point. 

Equivalently, for an actual quantum measurement to be completed, it is necessary for the branches corresponding to the different outcomes to decohere, and this implies that information is lost into the environment, hence the entropy relative to the observer increases.  See \cite{andrea} for a similar discussion in the context of the Relational Interpretation and \cite{Bohmian} for the role of dissipation in Bohmian Mechanics'  measurement.   I shall not pursue a quantitative analysis of he quantum case here, leaving it to future work, but the above notes are sufficient to show that quantum measurement cannot produce a choice without raising entropy either. 

A  general observation underpins the fact that a choice has a thermodynamical cost.  Thermal fluctuations are not predictable because they refer to the dynamics of the microscopic degrees of freedom and these are defined by their inaccessibility.  If we {\em measure} a specific thermal or quantum fluctuation, we  get a way to extract novel information from a thermal or quantum system; but {\em any measurement implies dissipation}.   The same physical principle underpins the thermodynamical cost of learning \cite{Milburn}.

The first to notice the basic and --seems to me-- under-appreciated fact that any measurement is dissipative is, as far as I know, Reichenbach in \cite{Reichenbach}. In turn, the fact that measurement implies dissipation follows from an even simpler observation: a measurement records information {\em in the future} of the measurement interaction.  This breaks time reversal invariance, and the only physical source for the breaking of time reversal invariance is thermodynamical irreversibility, namely dissipation. 

In summary, choosing can be modelled as bringing information from the micro (or quantum) physics to the macro (and classical) world.   This process necessitates to be stabilized, and this can only happen via dissipation. The overall entropic balance is necessarily positive: the information gained in the choice is over-compensated by what is lost in the associated dissipation.  The second law is not circumvented.  As the Germans say `Wahl ist Leiden', to choose is to suffer: perhaps not sufferance, but certainly dissipation.

\vspace{1cm}

\centerline{***}

\vspace{.2cm}

This work was made possible through the support of  the ID\#62312  grant from the John Templeton Foundation, as part of the ``The Quantum Information Structure of Spacetime (QISS)'' Project (\href{qiss.fr}{qiss.fr}). I thank all the participant to the August 2023 FQXi retreat, where I have found key ideas for completing this note. 
\vfill


\providecommand{\href}[2]{#2}\begingroup\raggedright\endgroup

\end{document}